\def\e{\varepsilon}
\def\l{\hat{l}}
\def\lp{\hat{l}'}
\def\lo{\hat{l}_0}
\def\lop{\hat{l}_0'}
\def\lolop{\lo \cdot \lop}
\def\n{n(\l,\e)}
\def\np{n(\lp,\e)}
\def\lu{\l \cdot \vec{u}}
\def\lpu{\lp \cdot \vec{u}}
\def\lou{\lo \cdot \vec{u}}
\def\lopu{\lop \cdot \vec{u}}
\def\lub{\left(\l \cdot \vec{V}\right)}
\def\nj{n}
\def\nf{n^i}
\def\npr{n^{ij}}
\def\nq{n^{ijk}}
\def\Hr{H^i}
\def\Pr{K^{ij}}
\def\Qr{Q^{ijk}}
\def\Vb{\vec{V}}
\documentstyle[12pt,aaspp4]{article}
 
\lefthead{Psaltis \& Lamb}
\righthead{Compton Scattering by Static and Moving Media}

\begin{document}

\begin{center}
To appear in {\em The Astrophysical Journal}
\end{center}

\title{Compton Scattering by Static and Moving Media\\
I. The Transfer Equation and Its Moments}
\vspace{0.5cm}
\author{Dimitrios Psaltis and Frederick K. Lamb}
\vspace{0.5cm}
\affil{Physics and Astronomy Departments,
       University of Illinois at
       Urbana-Champaign\altaffilmark{1}\\
      demetris@astro.uiuc.edu and f-lamb@uiuc.edu}
\altaffiltext{1}{Postal address: 1110 W. Green St., Urbana, IL 61801-3080,
USA}
 
\begin{abstract}
Compton scattering of photons by nonrelativistic particles is
thought to play an important role in forming the radiation
spectrum of many astrophysical systems. Here we derive the
time-dependent photon kinetic equation that describes
spontaneous and induced Compton scattering  as well as
absorption and emission by static and moving media, the
corresponding radiative transfer equation, and their zeroth
and first moments, in both the system frame and in the frame
comoving with the medium. We show that it is necessary to use
the correct relativistic differential scattering cross
section in order to obtain a photon kinetic equation that is
correct to first order in $\e/m_e$, $T_e/m_e$, and $V$, where
$\e$ is the photon energy, $T_e$ and $m_e$ are the electron temperature and rest
mass, and $V$ is the electron bulk velocity in units of the
speed of light. We also demonstrate that the terms in the
radiative transfer equation that are second-order in $V$
usually should be retained, because if the radiation energy
density is sufficiently large compared to the radiation flux,
the effects of bulk Comptonization described by the terms that
are second-order in $V$ are at least as important as the
effects described by the terms that are first-order in
$V$, even when $V$ is small. The system- and fluid-frame
equations that we derive are correct to first order in
$\e/m_e$. Our system-frame equations, which are correct to
second order in $V$, may be used when $V$ is not too large.
Our fluid-frame equations, which are exact in $V$, may be
used when $V \rightarrow 1$. Both sets of equations are valid
for systems of arbitrary optical depth and can therefore be
used in both the free-streaming and the diffusion regimes. We
demonstrate that Comptonization by the electron bulk motion
occurs whether or not the radiation 
\newpage \noindent
field is isotropic or the bulk flow converges and that it is
more important than thermal Comptonization if $V^2 >
3T_e/m_e$.

\end{abstract}

\keywords{plasmas -- radiation mechanisms: thermal -- radiation
transfer}

\section{INTRODUCTION}

Compton scattering of photons by nonrelativistic particles is
thought to play an important role in many astrophysical
settings, including the early universe (see
Peebles\markcite{P1971} 1971; Sunyaev
\& Zeldovich\markcite{SZ1980} 1980), clusters of galaxies (see
Rephaeli\markcite{r1995} 1995), active galactic nuclei (see
Mushotzky, Done, \& Pounds\markcite{DP1993} 1993), compact
galactic X- and gamma-ray sources (see Pozdnyakov, Sobol,
\& Sunyaev\markcite{PSS1983} 1983), and supernova remnants
(see McCray\markcite{MC1993} 1993). In this process photons
lose energy to the electrons or gain energy from
their thermal and bulk motions, as a result of
Compton recoil.          

Starting from the Boltzmann equation for photons,
Kompaneets\markcite{K1957} (1957) derived the so-called
{\em Kompaneets equation}, which describes the time
evolution of the photon energy distribution caused by
scattering by thermal electrons when there is no bulk
motion, the radiation field is perfectly isotropic, and
the change in the energy of the photon in each
scattering is small. The conditions assumed in deriving the
Kompaneets equation are never strictly satisfied in
astrophysical systems, since the radiation field is always
anisotropic near their boundaries. In inhomogeneous systems,
the radiation field may be anisotropic even in the interior.
Also, in many astrophysical systems the scattering
particles have substantial bulk motions. Thus, although
the Kompaneets equation has been used extensively to
treat astrophysical systems, in many cases it does not
give accurate results. This has motivated several authors
to derive photon kinetic equations that are valid under more
general conditions. 

Comptonization in regions where the radiation field is
anisotropic has been studied either by using Monte Carlo
techniques  or by solving partial differential equations
for the specific intensity of the radiation field using
finite-difference methods. Pozdnyakov et al.\
(1983) have reviewed Monte Carlo methods and results.
Nagirner \& Poutanen\markcite{NP1994} (1994) have
reviewed work based on calculation of the complete
Compton scattering kernel for polarized radiation.
Babuel-Peyrissac \& Rouvillois\markcite{BPR1969} (1969),
Pomraning\markcite{P1973} (1973), Payne\markcite{P1980}
(1980),  Madej\markcite{M1989}\markcite{M1991} (1989,
1991), and Titarchuk\markcite{T1994} (1994) derived 
radiative transfer equations that describe Comptonization
of an anisotropic radiation field when there is no bulk
motion. Chan \& Jones\markcite{CJ1975} (1975), Blandford \&
Payne\markcite{BP1981a} (1981a), and Fukue, Kato, \&
Matsumoto\markcite{FKM1985} (1985) derived photon kinetic
equations for thermal particles with non-zero bulk velocity,
in the diffusion approximation. Thorne\markcite{T1981}
(1981; see also Thorne, Flammang, \&
$\dot{\mbox{Z}}$ytkow\markcite{TFZ1981} 1981) derived 
fluid-frame moments of the radiative transfer equation in
general relativity. The equations derived by these various
sets of authors have been widely used to study
Comptonization by strong shocks and accretion
flows onto compact objects (see, e.g., Blandford \&
Payne\markcite{BP1981b} 1981b; Payne \&
Blandford\markcite{BP1981} 1981; Lyubarskij \&
Sunyaev\markcite{LS1982} 1982; Colpi\markcite{C1988} 1988;
Riffert\markcite{R1988} 1988; Mastichiadis \&
Kylafis\markcite{MK1992} 1992; Miller \&
Lamb\markcite{ML1992} 1992; Titarchuk \&
Lyubarskij\markcite{TL1995} 1995; Turolla et
al.\markcite{TZZN1996} 1997; Titarchuk, Mastichiadis, \&
Kylafis\markcite{TMK1997} 1997).

In the course of our investigation of the effects of
Comptonization on the X-ray spectra of accreting neutron
stars (see, e.g., Lamb\markcite{L1989} 1989; Miller \&
Lamb\markcite{ML1992} 1992; Psaltis, Lamb, \&
Miller\markcite{PLM1995} 1995) we have rederived the
radiative transfer equation and its moments for static and
moving media and found important corrections to almost all
the above derivations of the photon kinetic or radiative
transfer equations, as we explain in \S~2. There we show
that it is necessary to use the correct relativistic
differential scattering cross section in order to obtain a
photon kinetic equation that is correct to first order in
$\e/m_e$, $T_e/m_e$, and $V$, where $\e$ is the photon
energy, $T_e$ and $m_e$ are the electron temperature and
rest mass, and $V$ is the electron bulk velocity in units
of the speed of light (we use units in which the Boltzmann
constant and the speed of light are equal to unity). In
\S~2 we also demonstrate that the terms in the radiative
transfer equation that are second-order in $V$ usually
should be retained, because in many situations the
second-order terms are at least as important as the
first-order terms, even when $V$ is small (see also Yin
\& Miller 1995). If the terms that are second-order in
$V$ are instead neglected, significant errors are introduced
in the photon kinetic equation and its moments.

In \S~3 we state our assumptions and approximations and
introduce our notation. In \S~4 we derive the
time-dependent photon kinetic equation that describes
spontaneous and induced scattering by static and moving
media, the corresponding radiative transfer equation, and
their zeroth and first angular moments, in the system
frame and in the frame comoving with the medium. We
derive the moment equations as well as the kinetic and
transfer equations because, although it is usually
necessary to solve the full radiative transfer equation
in order to determine accurately the angular distribution
of the radiation field, the moment equations can be used
to speed up the numerical computation by a large factor
(Mihalas\markcite{M1980} 1980; see also Mihalas 1978,
pp.~157--158) and to provide additional physical insight.
The system- and fluid-frame equations that we derive are
correct to first order in $\e/m_e$. Our system-frame
equations, which are correct to second order in $V$, may
be used when $V$ is not too large. Our fluid-frame
equations, which are exact in $V$, may be used when $V
\rightarrow 1$. Both sets of equations are valid for
systems of arbitrary optical depth and can therefore be
used in both the free-streaming and the diffusion
regimes. Our equations can easily be generalized to
describe scattering by an arbitrary number of particle
species. 

In Appendix A we give the photon kinetic and radiative
transfer equations that are obtained by averaging the
equations for a single electron over a drifting, relativistic
Maxwellian electron velocity distribution. In Appendix B we
give the radiative transfer equation that describes
absorption and emission in moving media, and its zeroth and
first moments. There we point out that the addition of a
photon source term in the transfer equation without any
corresponding absorption term (see, e.g., Blandford \& Payne
1981) is fundamentally inconsistent with thermodynamics and
leads to a radiative transfer equation that has a different
mathematical character from the thermodynamically
consistent equation.

Finally, in \S~5 we summarize our results and their
implications for Comptonization by static and moving
media.

\section{MOTIVATION}

The radiative transfer equation that describes
scattering of photons by particles is an
integro-differential equation in which only derivatives
with respect to the spatial coordinates appear (see,
e.g., Nagirner \& Poutanen 1994). The scattering kernel
in this equation is non-local in photon energy and
depends on the (possibly complicated) correlations
between the angular dependence of the specific intensity
of the radiation field, the velocity distribution of the
particles, and the differential scattering cross
section. In order to accelerate numerical calculations,
gain better physical insight, and facilitate comparison
with previous studies that made similar approximations,
we convert this integro-differential equation into a
partial differential equation over the spatial
coordinates and photon energy by expanding the
scattering kernel in powers of the dimensionless
quantities $\e/m_e$, $T_e/m_e$, and $V$, which we assume
are small compared to unity. In this way, the scattering
kernel in the transfer equation becomes local in photon
energy and the scattering integral over solid angle can
be expressed in terms of the angular moments of the
specific intensity of the radiation field.

We are primarily interested in deriving a transfer
equation that can be used to calculate the spectra of
X-ray and soft $\gamma$-ray sources, so in expanding the
scattering kernel we shall keep only terms of zeroth and
first order in ${\Delta\e_0}/{\e_0} \approx {\e}/{m_e}
\ll 1$, which is the average fractional decrease in the
energy of a photon in a single scattering in the electron
rest frame (see eq.~[\ref{ScatteredEnergy}] below).
Then, in order to obtain a radiative transfer equation of
consistent accuracy, the terms in the expansion of the
kernel in powers of $T_e/m_e$ and $V$ that are of the
same size as the terms of first order in
${\Delta\e_0}/{\e_0}$, i.e., of the same size as
${\e}/{m_e}$, must be retained. Depending on the
situation, terms of different order in $T_e/m_e$ and $V$
may need to be included.

We now discuss the accuracy of the expansion of the
differential scattering cross section needed to
obtain a transfer equation that is accurate to first
order in $\e/m_e$, the orders of the terms in $V$ that
must be retained, and some subtle points that must be
taken into account if the diffusion approximation is used.

\subsection{Approximate Scattering Cross Section}

In order to obtain a transfer equation that is
consistently accurate to first order in $\e/m_e$, it is
necessary to use the correct relativistic expression for
the differential scattering cross section in the frame in
which one is working. We shall work in the electron rest
frame and we therefore use the Klein-Nishina expression
for the differential scattering cross section (see
eq.~[\ref{KN}]). To first order in $\e/m_e$, the
Klein-Nishina cross section is
 \begin{equation}
   \frac{d\sigma}{d\Omega_0}\approx
   \frac{3\sigma_T}{16 \pi}\Biggl[1+(\lolop)^2\Biggr]
   \Biggl [1-2\frac{\e}{m_e}
   (1-\lolop)\Biggr ]\;,
 \label{DifferentialCrossSection}
 \end{equation}
where $\sigma_{T}$ is the total cross section for Thomson
scattering and $\lo$ and $\lop$ are the direction vectors
of the photon in the electron rest frame, before and after
the scattering.

Use of the Thomson approximation for the differential
scattering cross section (Chan \& Jones 1975; Payne 1980;
Madej 1989, 1991) rather than
expression~(\ref{DifferentialCrossSection}) introduces
errors in the radiative transfer and moment equations of
order $\e/m_e$, i.e., of the same size as the basic Compton
effect. Neglecting the angle dependence of the term of order
$\e/m_e$ in expression~(\ref{DifferentialCrossSection})
(Pomraning 1973; Blandford \& Payne 1981a) introduces errors
in the radiative transfer equation of this same order because,
unlike the Thomson approximation,
expression~(\ref{DifferentialCrossSection}) is not
forward-backward symmetric. Approximating
expression~(\ref{DifferentialCrossSection}) by the
average of the cross section over the electron velocity
distribution (Titarchuk 1994) neglects the effects of
correlations between the angle dependence of the
differential cross section, the specific intensity, and
the electron velocity, introducing errors in the
radiative transfer equation of order $\e/m_e$,
$T_e/m_e$, and $V$. Finally, use in the system frame of
the Thomson or Klein-Nishina expressions for the
differential scattering cross section (Pomraning 1973;
Payne 1980; Madej 1989, 1991) introduces errors in the
transfer equation of order $V$ and $T_e/m_e$, because these
expressions are valid only in the electron rest frame.

\subsection{Importance of the Terms of Order $V^2$}

It has long been recognized that if the divergence of the
electron bulk velocity is non-zero, the electron bulk
motion causes a secular change in the energy of the
photons (see, e.g., Chan \& Jones 1975; Blandford \&
Payne 1981a; Fukue et al.\ 1985). The photon
kinetic equations derived in these works do not include
any terms of second order in $V$ and predict that photons
are systematically upscattered in energy by the electron
bulk motion if the flow is converging. However, as we
show with the examples that follow, photons are
systematically upscattered by the electron bulk motion
even if the terms that are first-order in $V$ have, on
average, no effect. Indeed, under some circumstances
photons are systematically upscattered by the bulk motion
even if the flow is {\it diverging}.

To see this, consider a situation in which the bulk
motion of the electrons can be described as isotropic
turbulence in which the velocity correlation length is
much smaller than the photon mean free path and the bulk
velocity satisfies $\nabla \cdot \Vb=0$. If all the terms
that are first-order in $V$ are included correctly in the
photon kinetic equation but terms that are second-order
in $V$ are not included (Fukue et al.\ 1985), the zeroth
and first moments of the kinetic equation will indicate
that the electron bulk motion has no effect on the photon
energy distribution on time scales longer than the
velocity correlation time. In reality, however, under the
conditions stated the effect of the electron turbulent
motions is completely analogous to the effect of electron
thermal motions, that is, the turbulent motions cause the
mean energy of the photons to increase steadily with time,
at a rate proportional to the second moment of the
turbulent velocity field. Hence, the electron bulk motion
causes the mean photon energy to increase with time even
though the flow is not converging; in fact, in this case
no terms that are first-order in $V$ have any effect on
the radiation field, on average. If the turbulent velocity
is high enough, the terms that are second-order in $V$
will cause the mean photon energy to increase even in a
diverging flow.

{\it Origin of terms second-order in $V$}.---The origin
of the terms in the photon kinetic equation that are
second-order in $V$ can be understood by considering a cold
($T_e = 0$) electron fluid with a uniform bulk velocity $\Vb$
in the system frame. For simplicity, let us analyze
scattering of photons in the Thomson limit in the electron
rest frame. In this limit, the angular distribution of the
scattered photons is backward-forward symmetric in the
electron rest frame, so the average photon energy in the
system frame after scattering is $\gamma^2\,(1-\langle{\hat
k}\cdot\Vb\rangle)$ times the energy before the
scattering (see Rybicki \& Lightman 1979, p.~198), where
$\gamma \equiv (1-V^2)^{-1/2}$ and the average is over
the initial photon wave vectors $\vec k$ in the system
frame. Consider first the case of a photon distribution
that is isotropic in the system frame. In this case the
average photon energy in the system frame after
scattering is $\gamma^2$ times the energy before
scattering, so in each scattering the average energy of a
photon is increased by an amount of order $V^2$ and hence
there must be terms of order $V^2$ in the kinetic
equation that describe this effect. In the more general
case of a photon distribution that is anisotropic in the
system frame these terms of order $V^2$ almost always
produce a secular increase in the average energy of the
photons, but there are also terms of order $V$ in the
kinetic equation that may cause the average energy of the
photons to increase or decrease, depending on the
velocity field and the angular distribution of the
photons (there are generally terms of order $V$ even if
$\nabla \cdot \Vb=0$; see Fukue et al.\ 1985).

{\it Relative sizes of terms of different order in
$V$}.---The order in $V$ of a given term in the various
moment equations does not by itself determine whether the
term should be retained in solving a given transport
problem. This is because the terms of different order in
$V$ involve different angular moments of the radiation
field, so the relative sizes of the moments must also
be considered in determining which terms should be
retained (see also Yin \& Miller 1995).

For example, in the zeroth moment of the radiative
transfer equation, which is a scalar equation for the
energy density of the radiation field, the terms that are
first-order in $V$ involve the scalar product of the
vector quantity $\vec{V}$ with the vector quantity
defined by the radiation field, i.e., the radiation flux
$\vec{H}$ (see eq.~[\ref{0thmoment}]). In contrast, the
terms that are second-order in $V$ do not involve
$\vec{H}$ but instead involve the scalar
quantity defined by the radiation field, i.e., the
radiation energy density $J$, and the radiation
stress-energy tensor $K^{ij}$ (again see
eq.~[\ref{0thmoment}]). In general, $V_iV_jK^{ij}$ is of
the same size as $V^2J$. Hence, if $J$ is sufficiently
large compared to $H$, the terms in the zeroth moment of
the radiative transfer equation that are second-order in
$V$ are generally at least as large as the terms that are
first-order in $V$.

In the first moment of the radiative transfer equation,
which is a vector equation for the radiation flux, the
situation is reversed, in that the terms that are
first-order in $V$ involve $J$ and $K^{ij}$ (see
eq.~[\ref{1stmoment}]) whereas the terms that are
second-order in $V$ involve
$\vec{H}$ and $Q^{ijk}$. In general, $V_jV_kQ^{ijk}$
is of the same size as $\vec{V} (\vec{V} \cdot \vec{H})$.
Hence, one might be tempted to neglect the terms of order
$V^2$ in the first moment of the radiative transfer
equation if $J$ is large compared to $H$, even if the
terms of order $V^2$ are retained in the zeroth moment
equation. However, this is in general unsafe, because it
involves treating the same terms in the radiative transfer
equation, from which the moment equations are derived,
differently, in taking the zeroth and first moment. In
addressing a given transport problem, one can only
determine which terms in the expansion in powers of $V$
must be kept by considering the boundary conditions as
well as the transfer equation.

The radiation field quantities that appear in terms that
involve only {\it odd\/} powers of $V$ are all of
about the same size. Similarly, the radiation field
quantities that appear in terms that involve only {\it
even\/} powers of $V$ are all of about the same size,
although they may be much larger or smaller than the
radiation field quantities that appear in the terms
that involve odd powers of $V$. Hence, when $V$ is $\ll
1$, terms of order $V^3$ and higher may be safely
neglected in the derivation of the transfer equation (see
also Yin \& Miller 1995).

\subsection{Use of the Diffusion Approximation}

In situations in which the change during a
scattering of the energy of a photon as measured in the
fluid frame can be neglected and the
longest photon mean free path
$\lambda_{\rm max}$ is much smaller than the smallest
length scale $L_{\rm min}$ on which physical variables
change, the specific intensity
in the frame comoving with the electron fluid,
$I_f(\e)$, is given to lowest order in $\lambda_{\rm
max}/L_{\rm min}$ by (see Mihalas \& Mihalas 1984,
p.~457)
 \begin{equation} 
 I_f(\e) = J_f(\e) + 3\,\l\cdot\vec{H}_f(\e)\;,
 \label{DiffusionApprox} 
 \end{equation}
 where $J_f(\e)$ and $\vec{H}_f(\e)$ are the zeroth and
first angular moments of $I_f(\e)$. The diffusion
approximation (sometimes called the zeroth-order
diffusion approximation; see Mihalas \& Mihalas 1984,
\S~97) consists in {\it assuming\/} that $I_f(\e)$ is
given {\it exactly\/} by
equation~(\ref{DiffusionApprox}). Then the first
fluid-frame Eddington factor $f^{ij}_f(\e) \equiv
K^{ij}_f(\e)/J_f(\e)$ is {\it exactly\/} equal to
$\small{1\over3}$; here $K^{ij}_f(\e)$ is the second moment
of $I_f(\e)$. As is evident from
equation~(\ref{DiffusionApprox}), if the source function
involves only the zeroth, first, and second moments of
$I_f(\e)$, then in the diffusion approximation it is
only necessary to solve the three equations
consisting of the zeroth and first moments of the fluid-frame
radiative transfer equation and the closure relation
$f^{ij}_f(\e) = \small{1\over3}$ for the three moments
$J_f(\e)$, $\vec{H}_f(\e)$, and $K^{ij}_f(\e)$; the specific
intensity may then be calculated in this approximation using
equation~(\ref{DiffusionApprox}).

The moments of the specific intensity in the system
frame do {\it not\/} satisfy
equation~(\ref{DiffusionApprox}), even when
$\lambda_{\rm max}/L_{\rm min} \ll 1$. If
expression~(\ref{DiffusionApprox}) is mistakenly used to
relate the specific intensity to its zeroth and first moments
in the system frame (Blandford \& Payne 1981) rather than in
the fluid frame, then errors of order $V$ will be introduced
in the radiative transfer equation and its moments, even if
$\lambda_{\rm max}/L_{\rm min} \ll 1$ (see Fukue et al.\
1985).

If one wants to use the diffusion
approximation~(\ref{DiffusionApprox}) to solve for the
moments of the specific intensity {\it in the system
frame}, one must solve the {\it four\/} equations
consisting of the zeroth and first moments of the
system-frame radiative transfer equation and the two
closure relations involving the first two Eddington
factors for the first four moments $J(\e)$, $H^i(\e)$,
$K^{ij}(\e)$, and $Q^{ijk}(\e)$ of the specific
intensity $I(\e)$ in the system frame. The reason is
that when equation~(\ref{DiffusionApprox}) is boosted
into the system frame and the terms that are second-order in
$V$ are retained, as in general they must be (see
above), the third moment of the specific intensity
$Q^{ijk}(\e)$ is introduced into the closure relation.
(Fukue et al.\ [1985] were able to set up a closed
system of equations consisting of the zeroth and first
moments of the system-frame radiative transfer equation
and a closure relation involving the first Eddington
factor only because they neglected all the terms in the
radiative transfer equation that are second-order and higher
in $V$.)

When, as in Compton scattering, the
energy of a photon in the fluid frame changes in a
scattering, expression~(\ref{DiffusionApprox}) may not be
accurate for all photon energies, even if the mean free path
of a photon is independent of its energy and
$\lambda_{\rm max}/L_{\rm min} \ll 1$. As an example,
consider a slab that is infinite along the $x$ and $y$ axes
but finite along the $z$ axis, in which the electrons are
cold and have no bulk motion. Suppose the slab is
illuminated from one side with monochromatic photons of
energy $\e_{\rm in}$ propagating in the $z$ direction. Since
the electrons are static, any photons that have scattered
have lost energy to the electrons and hence no longer have
energy $\e_{\rm in}$. Therefore all photons anywhere in the
slab with energies equal to $\e_{\rm in}$ have never been
scattered and are still propagating in the $z$
direction. As a result, the angular distribution of these
photons is not described accurately
by equation~(\ref{DiffusionApprox}), even if
$\lambda_{\rm max}/L_{\rm min}$ is very small; indeed,
{\it the Eddington factor at energy $\e_{\rm in}$ is equal to
unity everywhere in the slab\/}. In this example,
expression~(\ref{DiffusionApprox}) is accurate only at
energies sufficiently below the injection energy
$\e_{\rm in}$.

The electron bulk velocity $\vec{V}$ is a vector
quantity, and hence the relative sizes of the terms of
first and higher order in $V$ in the expansion of the
transfer equation and its moments depend strongly on the
relation between the angular dependence of the bulk
velocity, the radiation field, and the differential
scattering cross section. Therefore, even when the
diffusion approximation is only slightly inaccurate (as
for example when the Eddington factor differs only
slightly from $\small{1\over3}$), this inaccuracy
produces additional terms in the moment equations that
are of second order in $V$ and that are therefore of the
same magnitude as the second-order terms that would be
present if the radiation field were given exactly by
equation~(\ref{DiffusionApprox}). 

For these reasons we derive the radiative transfer and
moment equations without making use of the diffusion
approximation.

\section{ASSUMPTIONS, DEFINITIONS, AND APPROXIMATIONS}

In the sections that follow we assume that the electron
gas is nondegenerate (electron occupation number $\ll
1$). For conciseness we shall refer to it as a fluid,
without implying anything about whether it is collisional
or collisionless. We assume that photons are scattered
only by electrons, neglecting scattering by any other
particles, and set $h=c=k_B=1$, where
$h$ is Planck's constant, $c$ is the speed of light, and
$k_B$ is Boltzmann's constant. We indicate quantities
evaluated in the rest frame of a particular electron by a
subscript `0' and quantities evaluated in the inertial
frame momentarily comoving (locally) with the fluid,
which we call the `fluid frame', by a subscript `f'. The
`system frame' may be any global inertial frame (such as the
frame at rest with respect to the center of mass of the
system, if it is inertial). Quantities evaluated in the system
frame have no subscript.

{\it Bulk velocity and temperature}.---We define the
fluid frame as the frame in which the energy and momentum
flux density both vanish (Landau \&
Lifshitz\markcite{LL1987} 1987, p.~505). In this frame
collisions between the electrons tend to establish an
isotropic velocity distribution.

The system-frame three-velocity $\vec u$ of a given
electron is related to its fluid-frame three-velocity
${\vec u}_f$ by a Lorentz boost. The first and
second moments of $\vec u$ in the system frame are
 \begin{equation}
 \langle\vec{u}\rangle = \Vb
 \label{u1st}
 \end{equation}
and
 \begin{equation}
 \langle u^2 \rangle = \langle v^2 \rangle + V^2\;,
 \label{u2nd}
 \end{equation}
 where $\vec V$ is the three-velocity of the fluid as
measured in the system frame and
 \begin{equation}
 \vec{v} \equiv \vec{u}-\Vb
 \label{Defv}
 \end{equation}
 is the peculiar velocity of an electron as measured
in the system frame. 

If the electron momentum distribution in the fluid frame
is a relativistic Maxwellian, then the second moment of
the electron velocity distribution evaluated in the fluid
frame is
 \begin{equation} 
 \langle u_f^2\rangle_f = \frac{3T_e}{m_e}
 + {\cal O}\left(\frac{T_e^2}{m_e^2}\right)
 \label{uf2nd}
 \end{equation}
 and the second moment of the electron peculiar velocity
distribution in the system frame is
 \begin{equation} 
 \langle v^2\rangle = \frac{3T_e}{m_e}
 + {\cal O}\left(\frac{T_e}{m_e}V^2\right)\;.
 \label{v2nd}
 \end{equation}

{\it Description of the radiation field}.---In section 4.1
and Appendix A we derive the equation that describes the
evolution of a radiation field interacting with a moving
electron fluid. We shall refer to this equation as the
{\em photon kinetic equation} when it is written in terms
of the photon mode occupation number and as the {\em
radiative transfer equation} when it is written in terms
of the specific intensity of the radiation field; the two
descriptions are equivalent (see, e.g.,
Mihalas\markcite{M1978} 1978, p.~32), but the radiative
transfer equation is more often used in astrophysical
problems. 

In deriving the photon kinetic equation we shall describe
the radiation field by the number $\n$ of photons with
energy $\e$ propagating in direction
$\hat{l}$ with a given polarization state (we suppress the
dependence on polarization state, because we consider only
unpolarized radiation). The first few moments of $\n$ are
 \begin{eqnarray}
 \nj & \equiv & \frac{1}{4\pi} \int \n \,d\Omega\;, 
   \label{nave}\\
 \nf & \equiv & \frac{1}{4\pi}\int \n\, l^i \,d\Omega\;,  \\
 \npr & \equiv & \frac{1}{4\pi}\int \n\, l^i l^j \,d\Omega\;,  \\
 \nq & \equiv & \frac{1}{4\pi}\int \n\, l^i l^j l^k \,d\Omega\;.
   \label{nijk}
 \end{eqnarray}
 In definitions (\ref{nave})--(\ref{nijk}) the dependence
of the moments on position and photon energy have been
suppressed for brevity. Here and below we display the
dependence of the photon occupation number on $\l$ and
$\e$ in order to distinguish it clearly from its zeroth
moment.

In writing the radiative transfer equation we shall
describe the radiation field by its specific intensity
$I(\l,\e)\equiv 2\e^3\n$; here the factor of two accounts
for the two photon polarization states. The first few
moments of $I(\l,\e)$ are
 \begin{eqnarray}
J & \equiv & \frac{1}{4\pi}\int I(\l,\e)\, d\Omega= 2\e^3 \nj\;, \\
H^i & \equiv & \frac{1}{4\pi}\int I(\l,\e)\, l^i\, d\Omega = 2\e^3 \nf\;,  \\
K^{ij} & \equiv & \frac{1}{4\pi}\int I(\l,\e)\, l^i l^j\,d\Omega = 2\e^3 \npr\;,
\\  
Q^{ijk} &\equiv& \frac{1}{4\pi}\int I(\l,\e)\, l^i l^j l^k\,d\Omega = 2\e^3
\nq\;,
 \end{eqnarray}
 where again we have suppressed the dependence of the
moments on position and photon energy.

{\it Compton scattering}.---In the rest frame of the
electron the differential cross section for scattering
of unpolarized radiation is (see Berestetski\v \i,
Lifshitz, \& Pitaevski\v \i\markcite{BLP1971} 1971, p.
297) 
 \begin{equation}
\frac{d\sigma}{d\Omega_0}=\frac{3\sigma_T}{16\pi} \left(
\frac{\e_0'}{\e_0}\right)^2\left[\frac{\e_0}{\e_0'}+\frac{\e_0'}
   {\e_0}-1+(\lolop)^2\right]\;,
  \label{KN}
 \end{equation}
 where $\sigma_T$ is the Thomson cross section, $\e_0$ and
$\lo$ are the energy and direction of the incident photon,
and
 \begin{equation}
 \e_o' =\frac{\e_0}{1-\frac{\e_0}{m_e}(1-\lolop)}
 \label{ScatteredEnergy}
\end{equation}
and $\lop$ are the energy and direction of the scattered
photon.

The energy and direction of propagation of a photon in
the electron rest frame are related to the same
quantities in the system frame by the Lorentz
transformations  
 \begin{equation}
\e_0 = \gamma \e(1-\lu) = \frac{\e}{\gamma(1+\lou)}
   \label{eo}
\end{equation}
and
\begin{equation}
\lo = \frac{\e}{\e_0}\left\{ \l +\left[\frac{\gamma-1}{u^2}
         \left(\lu\right)
         -\gamma \right]\vec{u}\right\}\;.
   \label{lo}
\end{equation}
The photon phase-space volume is Lorentz invariant (see
Mihalas 1978, p. 495), i.e.,
\begin{equation}
\e\,d\e\,d\Omega = \e_0\,d\e_0\,d\Omega_0\;,
\end{equation}
so
\begin{equation}
\n = n_0(\lo,\e_0)\;.
\end{equation}

{\it Validity of the approximations}.---In deriving the
photon kinetic and radiative transfer  equations and
their moments in the system frame, we shall retain terms
up to first order in
$\e/m_e$ and second order in $u$ (first order in
$T_e/m_e$), neglecting terms of order $(\e/m_e)u$ or
higher. These  equations are therefore valid when
 \begin{equation}
 \frac{\e}{m_e} \ll 1 \qquad {\rm and} \qquad
 \frac{\e}{m_e}\left(V^2+\frac{3T_e}{m_e}\right)^{1/2}
 \ll \left(V^2+\frac{3T_e}{m_e}\right) \ll 1\;.
 \end{equation}
 These conditions are usually satisfied in accretion onto
white dwarfs and neutron stars, but are not satisfied
in accretion onto black holes, because the bulk velocity
$V \rightarrow 1$ at the horizon.

Our expressions for the photon kinetic and radiative
transfer equations in the fluid frame are correct to all
orders in $V$ but only to first order in
$T_e/m_e$ and $\e_f/m_e$. These equations are therefore
valid when
 \begin{equation}
 \frac{\e_f}{m_e} \ll 1 \qquad {\rm and} \qquad
 \frac{\e_f}{m_e}\left(\frac{3 T_e}{m_e}\right)^{1/2}
 \ll \left(\frac{3 T_e}{m_e}\right) \ll 1
 \end{equation}
 and can be used where $V$ is $\simeq 1$.

\section{PHOTON KINETIC AND RADIATIVE TRANSFER EQUATIONS
\newline AND THEIR ZEROTH AND FIRST MOMENTS\ \ \ \ \ \ \ }

\subsection{Photon Kinetic Equation}

The photon kinetic equation in the system frame is
\begin{eqnarray}
k^\mu \partial_\mu n(\vec{k}) & = & \int \frac{d^3 k'}{(2\pi)^3} \int d^3p
    \left\{W\left(\vec{k}'\vec{p},\vec{k}\vec{p'}\right)
   n(\vec{k}')\left[1+n(\vec{k})\right]f(\vec{p})\left[1-f(\vec{p'})\right]
   \right.\nonumber\\
& & \;\;\;\;\;\;\;\;\;\;\;\;\;\;\;\;\;\;\;\;\;\;\;\;\;
   \left.- W\left(\vec{k}\vec{p},\vec{k}'\vec{p'}\right)
   n(\vec{k})\left[1+n(\vec{k}')\right]f(\vec{p})\left[1-f(\vec{p'})
   \right]\right\}\;,
   \label{mastereq}
\end{eqnarray}
where $k^\mu=(\e,\e\hat{l})$, $n(\vec{k})$ is the photon
occupation number, $\vec{p}$ is the electron momentum,
$f(\vec{p})$ is the electron momentum distribution,
$W\left(\vec{k}\vec{p},\vec{k}'\vec{p'}\right)$ is the
transition rate for the scattering $\vec{k}+\vec{p}
\rightarrow \vec{k}'+\vec{p'}$, and in writing the total
derivative on the left side we have used the Einstein
summation convention. The left side of equation
(\ref{mastereq}) is manifestly covariant. The
ride side is also covariant, if the appropriate transition
rates are used.

It is convenient to integrate first over the photon states
and then over the electron states, because the angular
integrals are then much simpler. We will therefore assume for
the moment that all the electrons have the same momentum
$\vec{p}$. Because each side of the photon kinetic equation
(\ref{mastereq}) is covariant, the left side can be evaluated
in the system frame and the right side in the electron
rest frame, at the wave vector $\vec{k}_0$ that corresponds
to the wave vector $\vec{k}$ of the photon in the system
frame. The resulting equation for electrons moving with
velocity ${\vec u} = {\vec p}/(\gamma m_e)$ is (compare with
Peebles 1971, p.~204)
 \begin{eqnarray}
  \left(\partial_t+l^i \partial_i\right)\e\n & = 
  & \e_0 n_{e0} \int d\Omega_0'
  \Biggl\{
     \left(\frac{\e_0'}{\e_0}\right )^{\!\!2}
         \frac{\partial \e_0'}{\partial \e_0}
     \left(\frac{d\sigma}{d\Omega'}\right)_0
         n_0(\lop,\e_0') \left[1+n_0(\lo,\e_0)\right]
     \nonumber\\
 & & \qquad\qquad\qquad
   - \left(\frac{d\sigma}{d\Omega'}\right)_0
      n_0(\lo,\e_0) \left[1+n_0(\lop,\e_0')\right]
 \Biggr\}\;,
  \label{starting_eq}
 \end{eqnarray}
 where $\partial_t\equiv\partial/\partial t$,
$\partial_i\equiv\partial/\partial x^i$, the $x^i$ are the
spatial coordinates, $n_{e0}$ is the electron density in the
electron rest frame, $\hat{l}_0$ and $\e_0$ are related to
$\hat{l}$ and $\e$ by equations (\ref{lo}) and (\ref{eo}),
and $\e_0'$ is related to $\e_0$ by
equation~(\ref{ScatteredEnergy}). The factor preceding the
cross section in the first term of the collision integral is
the Jacobian that corrects for the different phase spaces of
the incident and scattered photons.

We evaluate the collision integral in equation
(\ref{starting_eq}) by first using the Lorentz invariance of
the photon occupation number to relate
$n_0(\lop,\e_0')$ to $n(\hat{l}',\e')$ and relating
$n(\hat{l}',\e')$ to $n(\hat{l}',\e)$ by expanding 
$n(\hat{l}',\e')$ to second order in $\vec{u}$ and to first
order in $\e/m_e$, which gives (see Peebles 1971, p.~204)
\begin{eqnarray} 
n_0(\lop,\e_0') \simeq \np
% \nonumber \\ 
%&\qquad &
 + \left[ \left(\lop-\lo\right)\cdot \vec{u}
 +\left(\lou\right)\left(\lo-\lop\right)\cdot
 \vec{u}+\frac{\e}{m_e}\left(1-\lolop\right)\right]  \e \partial_\e\,\np
 \nonumber \\ 
 + \frac{1}{2}\left[\left(\lop-\lo\right)\cdot
 \vec{u}\right]^2
 \e^2\partial^2_\e\,\np \;,
 \qquad\qquad\qquad\qquad\qquad\qquad\qquad\quad
\end{eqnarray}
where $\partial_\e\equiv\partial/\partial \e$ and
$\partial_\e^2\equiv\partial^2/\partial \e^2$. We then use this result,
the Lorentz invariance of the photon distribution, and
$(d\sigma/d\Omega ')_0$ and $n_{e0}\e_0$ expanded to
second order in $\vec{u}$ to obtain the approximate
photon kinetic equation for electrons moving with velocity
${\vec u}$
 \begin{eqnarray}
 \left(\partial_t+l^i \partial_i\right)\n & = &
   \frac{3 n_e \sigma_T}{16\pi} \int
  d\Omega_0'\left[1+\left(\lolop\right)^2\right]
   \left[1-2\frac{\e}{m_e}\left(1-\lolop\right)\right] \cdot\nonumber\\
& &\qquad\qquad
  \left\{\left[1-\left(\lou\right)+\left(\lou\right)^2-u^2\right]
    \right.\nonumber\\
& &\qquad\qquad\quad +\left[\left(\lop-\lo\right)\cdot \vec{u}+2
  \left(\lou\right)\left(\lo-\lop\right)\cdot\vec{u}\right]
   \e \partial_\e \nonumber\\ 
& &\qquad\qquad\quad
 +\frac{1}{2}\left[\left(\lop-\lo\right)\cdot \vec{u}\right]^2
 \e^2 \partial^2_\e
   +\frac{\e}{m_e}\left(1-\lolop\right)\left(4+\e\partial_\e\right)
    \nonumber\\
& &\qquad\qquad\quad \left.+2 \frac{\e}{m_e} \left(1-\lolop\right) n(\l,\e)
   \left(2 + \e \partial_\e \right)\right\} \np \nonumber \\ 
& &\qquad\qquad\qquad\qquad  - n_e \sigma_T
   \left(1-2\frac{\e}{m_e}\right)\left(1-\lu\right) \n \;,
   \label{PKE}
\end{eqnarray}
which describes the effects of scattering by electrons with velocity
$\vec{u}$.

In Appendix A we give the photon kinetic and radiative
transfer equations that are obtained by averaging equation
(\ref{PKE}) over a drifting, relativistic Maxwellian
electron velocity distribution (see
eqs.~[\ref{u1st}]--[\ref{v2nd}]). The moment equations
derived in the next two subsections can be obtained by
computing the zeroth and first moments of the equations
given in Appendix A. Here we follow the simpler approach
of first computing the moments of equation (\ref{PKE})
and then averaging them over the electron velocity
distribution.

\subsection{Zeroth Moment and Radiation Energy Density}

We compute the zeroth moment of the photon kinetic equation by first
integrating both sides of equation (\ref{PKE}) over all directions.
Making use of the Jacobian 
\begin{eqnarray}
\frac{\partial(\Omega,\Omega_0')}{\partial(\Omega_0,\Omega ')} 
& =  & \left(\frac{\e_0}{\e}\frac{\partial\e_0}{\partial\e}\right)
  \left(\frac{\e '}{\e_0'}\frac{\partial\e '}{\partial\e_0'}\right)
  \nonumber\\ 
& \simeq &
  \left[1-2\left(\lou\right)+2\left(\lpu\right)-2u^2+3\left(\lou\right)^2
  \right.\nonumber\\
&  & \qquad\qquad\qquad\qquad
+3\left.\left(\lpu\right)^2-4\left(\lou\right)\left(\lpu\right)\right]\;,
   \label{jacobian}
\end{eqnarray} 
we find
\begin{eqnarray}
\partial_t \nj+\partial_i \nf & = &
   \frac{3 \sigma_T}{16\pi} n_e \int d\Omega ' 
   \int d\Omega_0 \left[1+\left(\lolop\right)^2\right]
\left[1-\frac{2\e}{m_e}\left(1-\lolop\right)\right] \cdot \nonumber\\ 
& &\Biggl\lbrack\!\!\Biggl\lbrack\Biggl\{
 \biggl[ 1-3u^2+3\left(\lpu\right)^2+2\left(\lpu\right) \biggr]
 +\left(\lopu\right)\left[1+ 2\left(\lpu\right)\right]\e\partial_\e
 \nonumber\\
& &\qquad\qquad\qquad\qquad\qquad\qquad\quad
 \! + \frac{1}{2}\left(\lopu\right)^2\e^2\partial^2_\e
       +\frac{\e}{m_e}\biggl(4+\e\partial_\e\biggr)\Biggr\}\nonumber\\
& &\quad - \left(\lou\right)\Biggl\{\left[3+6\left(\lpu\right)\right]+
   \left[1+2\left(\lpu\right)\right]\e\partial_\e
    +\left(\lopu\right)\e^2\partial^2_\e\Biggr\}
           \nonumber\\
& &\quad +\left(\lou\right)^2 \biggl( 6+4\e\partial_\e+
      \frac{1}{2}\e^2\partial^2_\e \biggr)
     - \frac{\e}{m_e}(\lolop) \biggl( 4 + \e\partial_\e \biggr)
       \nonumber\\
& &\qquad\qquad\qquad\qquad\qquad\qquad\qquad
 + \frac{2\e}{m_e} \left(1-\l\cdot\lp\right) n(\l,\e)
   \biggl( 2 + \e \partial_\e \biggr)
  \Biggr]\!\!\Biggr]\;\np\nonumber\\  
& &\qquad\qquad\qquad\qquad
- n_e\sigma_T \left[\left(1-2\frac{\e}{m_e}\right)\nj - \nf u_i\right]\;.
\end{eqnarray}
Next we integrate over $d\Omega_0$ and then over $d\Omega '$, using
definitions (\ref{nave})--(\ref{nijk}). The result is
\begin{eqnarray}
\frac{1}{ n_e \sigma_T}\left(\partial_t \nj+
  \partial_i\nf\right) & = & \left(3+\e\partial_\e\right) \nf u_i
   +\left[\frac{\e}{m_e}\left(4+\e\partial_\e\right)
   +\frac{u^2}{3}\left(4\e\partial_\e +
   \e^2\partial^2_\e \right)\right]\nj
   \nonumber \\
& & +\left(\frac{36}{10}+\frac{34}{10}\e\partial_\e
        +\frac{11}{20}\e^2\partial^2_\e\right)
        \left(\npr u_i u_j-\frac{1}{3}\nj u^2\right)\nonumber\\
& & +\frac{3}{2}
\left(\frac{\e}{m_e}\right)\left(2 n^2 - 2 \nf\nf + 2 \npr\npr 
  -2\nq \nq\right.\nonumber\\
& & \;\;\;\;\;\;\;\;\;\;\;\;\; +\left.n\e \partial_\e n-\nf\e \partial_\e \nf
 +\npr\e\partial_\e \npr-
    \nq\e\partial_\e \nq\right)\;,
   \label{0thmomentv1}
\end{eqnarray}
where we have again used the Einstein summation convention. Finally,
after averaging equation (\ref{0thmomentv1}) over the
electron velocity distribution we obtain the zeroth moment
of the kinetic equation that describes the effects of
scattering by a fluid of electrons, namely, 
\begin{eqnarray}
\frac{1}{ n_e \sigma_T}\left(\partial_t \nj+
  \partial_i\nf\right) & = & \left(3+\e\partial_\e\right) \nf V_i
   +\frac{\e}{m_e}\left(4+\e\partial_\e\right)\nj
   \nonumber \\
& & +\frac{1}{3}\left(\langle v^2\rangle +V^2\right)
   \left(4\e\partial_\e+\e^2\partial^2_\e\right)\nj
   \nonumber\\
& & +\left(\frac{36}{10}+\frac{34}{10}\e\partial_\e
        +\frac{11}{20}\e^2\partial_\e^2\right)
        \left(\npr\langle v_i v_j\rangle-
        \frac{1}{3}\nj\langle v^2 \rangle\right)
        \nonumber\\
& & +\left(\frac{36}{10}+\frac{34}{10}\e\partial_\e
       +\frac{11}{20}\e^2\partial^2_\e\right)
        \left(\npr V_i V_j-\frac{1}{3}\nj V^2\right)\nonumber\\
& & +\frac{3}{2}
   \left(\frac{\e}{m_e}\right)\left(2 n^2 - 2 \nf\nf + 2 \npr\npr 
  -2\nq \nq\right.\nonumber\\
& & \;\;\;\;\;\;\;\;\;\;\;\;\;+\left.n\e \partial_\e n-\nf\e \partial_\e \nf
 +\npr\e\partial_\e \npr-
    \nq\e\partial_\e \nq\right)\;.
   \label{0thmomentv2}
\end{eqnarray}
This equation is valid in both the diffusion and free-streaming regimes,
for any arbitrary (possibly anisotropic) distribution of electron
velocities.  

If the electron velocity distribution in the fluid frame is
a relativistic Maxwellian with temperature $T_e$, then (see
eqs.~[\ref{u1st}] and [\ref{v2nd}])
\begin{equation}
\langle v_i v_j \rangle 
\simeq \frac{1}{3}\langle v^2\rangle\delta_{ij} 
\simeq \frac{T_e}{m_e} \delta_{ij}\;,
\end{equation}
and the zeroth moment of the photon kinetic
equation can be written, to the same accuracy as equation
(\ref{0thmomentv2}), as
\begin{eqnarray}
\frac{1}{ n_e \sigma_T}\left(\partial_t \nj+
  \partial_i \nf\right) & = & \left(3+\e\partial_\e\right) \nf V_i
   +\left[\frac{\e}{m_e}\left(4+\e\partial_\e\right)
+\left(\frac{T_e}{m_e}+\frac{V^2}{3}\right)
   \left(4\e\partial_\e+
   \e^2\partial^2_\e\right)\right]\nj\nonumber\\
& & +\left(\frac{36}{10}+\frac{34}{10}\e\partial_\e+
      \frac{11}{20}\e^2\partial_\e^2\right)
        \left(\npr V_i V_j-\frac{1}{3}\nj V^2\right)\nonumber\\
& & +\frac{3}{2}
   \left(\frac{\e}{m_e}\right)\left(2 n^2 - 2 \nf\nf + 2 \npr\npr 
  -2\nq \nq\right.\nonumber\\
& & \;\;\;\;\;\;\;\;\;\;\;\;\;+\left.n\e \partial_\e n-\nf\e \partial_\e \nf
 +\npr\e\partial_\e \npr-\nq\e\partial_\e \nq\right)\;,
\end{eqnarray}
where we have used the relation $\npr \delta_{ij}=n$. The corresponding
zeroth moment of the radiative transfer equation with emission and
absorption included (see Appendix B) is
\begin{eqnarray}
\frac{\partial J}{\partial t}+ \frac{\partial}{\partial x^i}H^i & = &
  \left\{\e
  \partial_\e\left[\left(\frac{\e-4T_e}{m_e}\right)J\right]+
   \frac{T_e}{m_e}\e\partial_\e^2(\e J)
  + \e\partial_\e H^i V_i\right.
   \nonumber\\
& & \;\;\;\;\;\;\;\;\;\;\;\;\;
    +\frac{V^2}{3}\left[-4\e\partial_\e J+
   \e\partial^2_\e(\e J)\right]\nonumber\\
& & \;\;\;\;\;\;\;\;\;\;\;\;\;
     +\left(\frac{1}{10}\e\partial_\e+
     \frac{11}{20}\e^2\partial_\e^2\right)
     \left(K^{ij}V_i V_j
   -\frac{1}{3}J V^2\right) \nonumber\\ 
& & \;\;\;\;\;\;\;\;\;\;\;\;\;
   +\frac{3}{4}\left(\frac{\e}{m_e}\right)
   \left[\left(\e \partial_\e J - J\right)\frac{J}{\e^3}
    -\left(\e \partial_\e H^i - H^i\right)\frac{H^i}{\e^3}\right.
    \nonumber\\
& & \;\;\;\;\;\;\;\;\;\;\;\;\;\quad\quad\quad
   \left.\left.+\left(\e \partial_\e K^{ij} -
    K^{ij}\right)\frac{K^{ij}}{\e^3}
    -\left(\e \partial_\e Q^{ijk} -Q^{ijk}\right)\frac{Q^{ijk}}{\e^3}
     \right]\right\}n_e \sigma_T \nonumber\\
& & +\left(1-\frac{1}{6}V^2  
   \e\partial_\e+\frac{1}{6}V^2
   \e^2\partial_\e^2\right)\eta_\e\nonumber\\
& &-\left\{ \left( J+\frac{1}{2} J V^2 -H^i V_i\right) 
 +\left[K^{ij}V_i V_j+\frac{1}{2}JV^2 - H^i V_i
  \right]\e\partial_\e\right.\nonumber\\   
 & & \qquad\qquad\qquad\qquad\qquad\qquad\qquad\qquad
   \left.+\frac{1}{2} K^{ij}V_iV_j  \e^2
   \partial_\e^2\right\}\chi_\e\;,   
   \label{0thmoment}
\end{eqnarray}
where $\eta_\e$ and $\chi_\e$ are the emission and
absorption coefficients, which are defined in the fluid frame but evaluated
at the energy $\e$ of the photons in the system frame.

\subsection{First Moment and Radiation Flux}

We compute the first moment of the photon kinetic equation by multiplying
both sides of equation (\ref{PKE}) by $\l$ and then integrating over
all directions. In performing the integration we use
the transformation (\ref{lo}) and the Jacobian (\ref{jacobian}),
integrating first over
$d\Omega_0$ and then over $d\Omega'$. The result is
\begin{eqnarray}
\frac{1}{n_e \sigma_T}\left(\partial_t \nf 
 + \partial_j\npr \right) & = & - \nf
 -\frac{2}{5}\frac{\e}{m_e}\left(1 + \e  \partial_\e\right)\nf
 + \frac{1}{5}\left(- n u_i + 3 \npr u_j \right)
   \nonumber \\ 
 & & -\frac{1}{5}\left(8 \nq u_j u_k -2\nf u_j u_j\right)
   -\frac{1}{10}\left[3\e\partial_\e \left(\nj u_i\right)
   + \e\partial_\e \left(\npr u_j \right)
       \right] \nonumber \\
 & & -\frac{1}{10}\left[10 \e \partial_\e
   \left(n^j u_j\right)u_i + 9\e
   \partial_\e \left(\nq u_j u_k \right)
   - \e\partial_\e\nf u^2\right]\nonumber\\
 & & - \frac{1}{10}\left[3\e^2\partial_\e^2
   \left(n^j u_j\right)u_i
   +\e^2 \partial_\e^2 \left(\nq u_j u_k\right)\right]\nonumber\\
& &+\frac{3}{2}
   \left(\frac{\e}{m_e}\right)\left(2 n n^i - 2 n^j\npr 
   + 2 n^{jk} n^{ijk}  
   -2n^{jkl} n^{ijkl}\right.\nonumber\\
& & \;\;\;\;\;\;\;\;\;\;\;\;\;
   +\left.n^{i}\e\partial_\e n-\npr\e\partial_\e n^j
   +\nq\e\partial_\e n^{jk}-   
 n^{ijkl}\e\partial_\e n^{jkl}\right)\;.
  \label{1stmomentv1}
\end{eqnarray}
Averaging equation (\ref{1stmomentv1}) over the electron
velocity distribution, we finally obtain 
\begin{eqnarray}
\frac{1}{n_e \sigma_T}\left(\partial_t \nf 
 + \partial_j\npr \right) & = & - \nf
 -\frac{2}{5}\frac{\e}{m_e}\left(1 + \e \partial_\e\right)\nf
   \nonumber\\
& &+\frac{3}{5}\left(\npr V_j-\frac{1}{3}\nj V_i \right)
 -\frac{1}{10}\e\partial_\e
 \left(\npr V_j+3\nj V_i\right)\nonumber\\
& & -\frac{2}{5}\left(4 n^{ijk}V_jV_k-n^i V_j V_j\right)
  \nonumber\\
& & -\frac{1}{10} \e\partial_\e
   \left(9\nq V_j V_k+ 10 n^j V_j V_i 
   -n^i V_j V_j\right)\nonumber\\  
& & -\frac{1}{10}\e^2\partial_\e^2
   \left(\nq V_j V_k   +3n^j V_j V_i\right)\nonumber\\
& & -\frac{2}{5}\left(4 n^{ijk}
  \langle v_jv_k\rangle-n^i \langle v_j
  v_j\rangle\right)
  \nonumber\\
& & -\frac{1}{10} \e\partial_\e
   \left(\nq \langle v_j v_k\rangle
   +10 n^j \langle v_j v_i\rangle
   -n^i \langle v_j v_j\rangle\right)\nonumber\\  
& & -\frac{1}{10}\e^2\partial_\e^2
   \left(\nq \langle v_j v_k\rangle  +
   3n^j \langle v_j v_i \rangle\right)\nonumber\\
& &+\frac{3}{2}
   \left(\frac{\e}{m_e}\right)\left(2 n n^i - 2 n^j\npr 
   + 2 n^{jk} n^{ijk}  
   -2n^{jkl} n^{ijkl}\right.\nonumber\\
& & \;\;\;\;\;\;\;\;\;\;\;\;\;
   +\left.n^{i}\e\partial_\e n-\npr\e\partial_\e n^j
   +\nq\e\partial_\e n^{jk}-   
 n^{ijkl}\e\partial_\e n^{jkl}\right)\;.
\end{eqnarray}

If the electron velocity distribution in the fluid frame is
a relativistic Maxwellian with temperature $T_e$, then
\begin{equation}
n^j \langle  v_j v_i\rangle=\frac{T_e}{m_e} n^i
\end{equation}
and
\begin{equation}
\nq\langle v_j v_k\rangle = \frac{T_e}{m_e} \nf\;,
\end{equation}
and the first moment of the photon kinetic equation reduces to
\begin{eqnarray}
\frac{1}{n_e \sigma_T}\left(\partial_t \nf
 + \partial_j\npr \right) & = & -\left[1+
 \frac{2}{5}\frac{\e}{m_e}\left(1 + \partial_\e\right)
 +\frac{2}{5}\frac{T_e}{m_e}\left(1+4\e\partial_\e
 + \e^2 \partial_\e^2\right)\right]\nf
\nonumber\\
& &+\frac{3}{5}\left(\npr V_j-\frac{1}{3}\nj V_i \right)
 -\frac{1}{10}\e\partial_\e
 \left(\npr V_j+3\nj V_i\right)\nonumber\\
& & -\frac{2}{5}\left(4 n^{ijk}V_jV_k-n^i V_j V_j\right)
  \nonumber\\
& & -\frac{1}{10} \e\partial_\e
   \left(9\nq V_j V_k+ 10 n^j V_j V_i 
   -n^i V_j V_j\right)\nonumber\\  
& & -\frac{1}{10}\e^2\partial_\e^2
   \left(\nq V_j V_k   +3n^j V_j V_i\right)\nonumber\\
& &+\frac{3}{2}
   \left(\frac{\e}{m_e}\right)\left(2 n n^i - 2 n^j\npr 
   + 2 n^{jk} n^{ijk}  
   -2n^{jkl} n^{ijkl}\right.\nonumber\\
& & \;\;\;\;\;\;\;\;\;\;\;\;\;
   +\left.n^{i}\e\partial_\e n-\npr\e\partial_\e n^j
   +\nq\e\partial_\e n^{jk}-   
 n^{ijkl}\e\partial_\e n^{jkl}\right)\;.
\end{eqnarray}
The corresponding first moment of the radiative transfer equation with
absorption and emission included is (again see Appendix B)
\begin{eqnarray}
\partial_t\Hr+\partial_j\Pr & = &
 \Biggl\{-\Hr-\frac{2}{5}\left(\frac{T_e-3\e}{m_e}\right)\Hr-
 \frac{2}{5}\left[\e
\partial_\e\left(\frac{\e-4T_e}{m_e}\right)+\frac{T_e}{m_e}\e
   \partial_\e^2\e\right]\Hr\nonumber\\
& &+\frac{1}{10}
   \left[\left(9-\e\partial_\e\right)\Pr V_j
  +\left(7-3\e\partial_\e \right)J V_i\right] \nonumber\\
& &+\frac{1}{10}\left(-6+8\e \partial_\e
-\e^2\partial_\e^2\right) H^j V_j V_i
+\frac{1}{10}\left(1+\e \partial_\e\right) H^i V_j V_j
  \nonumber\\
& &-\frac{1}{10}\left(1+3\e\partial_\e
+\e^2\partial_\e^2\right)\Qr V_j V_k
 \nonumber\\
& & +\frac{3}{4}\left(\frac{\e}{m_e}\right)
    \left[\left(\e \partial_\e J - J\right)\frac{H^i}{\e^3}
    -\left(\e \partial_\e H^j - H^j\right)\frac{K^{ij}}{\e^3}\right.
    \nonumber\\
& & \;\;\;\;\;\;\;\;\;\;\;\;\;
     +\left.\left(\e \partial_\e K^{jk} -
     K^{jk}\right)\frac{Q^{ijk}}{\e^3}
    -\left(\e \partial_\e Q^{jkl} -Q^{jkl}\right)\frac{R^{ijkl}}{\e^3}
     \right]\Biggr\}n_e \sigma_T\nonumber\\
& & -\left[\left(1+\frac{1}{2}V_jV_j\right)H^i -K^{ij}V_j
   + \left(\frac{1}{2}H^i V_jV_j -K^{ij}V_j+
  \frac{1}{2}Q^{ijk}V_jV_k\right)\e\partial_\e\right.
 \nonumber\\
& & \qquad\qquad
 \left.+\frac{1}{2}Q^{ijk}V_jV_k\e^2\partial^2_\e
   \right]\chi_\e +\frac{1}{3} V_i\left(2-\e
   \partial_\e\right)\eta_\e\;,
  \label{1stmoment}
\end{eqnarray}
where $R^{ijkl}$ is the fourth moment of the specific intensity.

\subsection{Equations in the Fluid Frame}

The photon kinetic and radiative transfer equations as well
as their moments take their simplest form in the fluid
frame, since $V=0$ in this frame, by definition. These
equations can be written choosing as independent variables
either Eulerian coordinates fixed in space and time or
Lagrangian coordinates comoving with the fluid. 

In terms of the fluid-frame coordinates, the transfer
equation~(\ref{RTEave}) and its zeroth and first moment
(\ref{0thmoment}) and~(\ref{1stmoment}) in the fluid frame
become
 \begin{eqnarray}
 \left(\partial_{t_f}+l_f^i\partial_{i_f}\right)
     I_f(\l_f,\e_f)
 &=&  n_{e,f}\sigma_T \Biggl [ {\cal L}_1 I_f(\l_f,\e_f) 
  \nonumber\\
 & &\qquad + \frac{3}{4}  \left( {\cal
   L}_2 J_f  +{\cal L}_3^i\Hr_f + {\cal L}_4^{ij} \Pr_f +
   {\cal L}_5^{ijk} \Qr_f + {\cal L}_6^{ijkl}
   R^{ijkl}_f\right)\nonumber\\  
 & & \qquad 
+\frac{3}{4\e ^3}\left(\frac{\e_f}{m_e}\right)I_f(\l_f,\e_f)
  \left(\e_f \partial_{\e_f}-1\right)\nonumber\\
 & &\qquad\qquad   
   \left(J_f-l^i_f \Hr_f + l^i_f l^j_f \Pr_f - l^i_f l^j_f
  l^k_f \Qr_f\right) \Biggr]\;, 
   \label{RTEfluid}
 \end{eqnarray}
 \begin{eqnarray}
 \partial_{t_f} J_f+\partial_{i_f}H^i_f &=& n_{e,f} \sigma_T
 \Biggl\{\left[ \e_f\partial_{\e_f}
 \left(\frac{\e_f-4T_e}{m_e}\right)
 +\frac{T_e}{m_e}\e_f\partial_{\e_f}^2\e_f\right]J_f
  \nonumber\\
 & & \qquad\qquad
 +\frac{3}{4}\left(\frac{\e_f}{m_e}\right)
   \left[\left(\e_f \partial_{\e_f} J_f -
   J_f\right)\frac{J_f}{\e_f^3}
    -\left(\e_f \partial_{\e_f} H_f^i -
   H_f^i\right)\frac{H_f^i}{\e_f^3}\right.
    \nonumber\\
& & \;\;\;\;\;\;\;\;\;\;\;\;\;\quad \left.
   +\left(\e_f \partial_{\e_f} K_f^{ij} -
   K_f^{ij}\right)\frac{K_f^{ij}}{\e_f^3}
    -\left(\e_f \partial_{\e_f} Q_f^{ijk}
  -Q_f^{ijk}\right)\frac{Q_f^{ijk}}{\e_f^3}
     \right]\Biggr\}
  \label{0thmomentfluid}
 \end{eqnarray}
 and
 \begin{eqnarray}
 \partial_{t_f} H^i_f+\partial_{j_f}K^{ij}_f
 &=&-n_{e,f}\sigma_T\Biggl\{H^i_f+\frac{2}{5}\left[
 \left(\frac{T_e-3\e_f}{m_e}\right)
 +\e_f\partial_{\e_f}\left(\frac{\e_f-4T_e}{m_e}\right)
  +\frac{T_e}{m_e}\e_f
  \partial_{\e_f}^2\e_f\right]H^i_f\nonumber\\
& & \qquad\qquad+\frac{3}{4}\left(\frac{\e_f}{m_e}\right)
    \left[\left(\e_f \partial_{\e_f} J_f -
   J_f\right)\frac{H_f^i}{\e_f^3}
    -\left(\e_f \partial_{\e_f}  H_f^j -
   H_f^j\right)\frac{K_f^{ij}}{\e_f^3}\right.
    \nonumber\\
& & \;\;\;\;\;\;\;\;\;\;\;\;\;\qquad
     +\left.\left(\e_f \partial_{\e_f} K_f^{jk} -
     K_f^{jk}\right)\frac{Q_f^{ijk}}{\e_f^3}
    -\left(\e_f \partial_{\e_f} Q_f^{jkl}
    -Q_f^{jkl}\right)\frac{R_f^{ijkl}}{\e_f^3}
     \right]\Biggr\}\;,\nonumber\\
   \label{1stmomentfluid}
 \end{eqnarray}
 where the coefficients ${\cal L}_1$--${\cal L}_6$, which
are given in Appendix A, are to be evaluated at $V=0$. 

The left sides of the above equations can also be written in
terms of the system-frame Eulerian coordinates as in
equations (95.9), (95.11), and (95.12) of Mihalas \&
Mihalas (1984). The resulting equations, which are often
called {\it mixed-frame} equations, are correct to all
orders in $V$ and therefore can be used in situations where
the bulk velocity is relativistic. The velocity-dependent
terms that appear in the mixed-frame equations arise from
Lorentz transformation of the left sides of equations
(\ref{RTEfluid})--(\ref{1stmomentfluid}), whereas
the velocity-dependent terms
that appear in the system-frame equations arise from
Lorentz transformation of the scattering integral on the
right sides of these equations.

The right side of the radiative transfer
equation~(\ref{RTEfluid}), can also be used in the formalism
developed by Thorne (1981), for solving Comptonization
problems in general relativity (compare eqs.
[6.10] and [6.13] of Thorne 1981)

\section{DISCUSSION}

In the previous section we derived the radiative
transfer equation in both the system and fluid frames,
taking into account absorption and emission as well as
spontaneous and induced Compton scattering. In this
section, we first show that our equation reduces to the
Kompaneets equation in the appropriate limits and call
attention to several errors and misunderstandings in the
literature. Next, we discuss the moment equations for an
anisotropic radiation field in a static medium and show
that the radiation force, and hence the critical flux and
luminosity, generally depend on both the photon energy
spectrum and the electron temperature. We then consider the
equation for the zeroth moment of the specific intensity for
moving media and show that if the radiation field is
isotropic, the terms in the transfer equation that are
second-order in the electron bulk velocity produce a
systematic increase in the energy of the photons that is
completely analogous to the systematic increase in the
energy of the photons produced by the electron thermal
motions. We also show that Comptonization by electron bulk
motion occurs, whether or not the radiation field is
isotropic or the bulk flow converges, and give estimates for
the time scales on which the photon energy distribution
changes because of systematic downscattering and
upscattering caused by the electron thermal and bulk
motions. We derive a new, more general condition for
determining when Comptonization by the electron bulk motion
is more important than Comptonization by the electron
thermal motions. We conclude by indicating how the transfer
equations we have derived can be solved using the method of
variable Eddington factors.

\subsection{The Kompaneets Equation}

When the radiation field is isotropic and there is no
bulk velocity and no absorption or emission, equation
(\ref{0thmoment}) reduces to the Kompaneets (1957)
equation\footnote{
In their derivations of the Kompaneets equation, Rybicki
\& Lightman (1979, p.~213) and Katz (1987, pp.~100--114)
did not take into account in the collision integral the
different phase spaces of the incident and scattered
photons. However, following Kompaneets they evaluated
the integral by using photon conservation and the
thermodynamic equilibrium photon distribution rather
than by performing the integration directly and thereby
obtained the correct result despite this error. These
authors also used the Thomson approximation to the
Klein-Nishina cross section. In general this introduces
an error of the same size as the systematic
downscattering term (see below), but this error vanishes
if the photon distribution is isotropic.}
 \begin{equation}
 \partial_{t_c} J=
 \frac{\e}{m_e}\partial_\e(\e J)-
 \frac{4 T_e}{m_e}\e\partial_\e J +
 \frac{T_e}{m_e}\e\partial_\e^2(\e J) 
 + \frac{\e}{m_e}\left(\e\partial_\e J
   -J\right)\frac{J}{\e^3}\;,
  \label{Kompaneets}
 \end{equation}
 where on the left side we have introduced the differential
Compton time $dt_c\equiv n_e \sigma_T dt$. The first two
terms on the right side of equation (\ref{Kompaneets})
describe the effects of systematic downscattering and
upscattering of the photons by electrons. The third term
describes the diffusion in energy produced by the
thermal motion of the electrons. The last term describes the
effect of induced Compton scattering.

As noted by Kompaneets (1957), {\it any\/} Bose-Einstein
distribution is a stationary solution of
equation~(\ref{Kompaneets}). Induced Compton scattering
cannot change the chemical potential, because it does not
change the number of photons. Hence, the statement by
Pomraning (1973, p.~193; see also Rybicki \& Lightman
1979, p.~209) that the Planck spectrum (the particular
Bose-Einstein distribution with zero chemical potential)
is the only stationary solution of the Kompaneets
equation if induced scattering is included is not
correct. Integrating equation~(\ref{Kompaneets}) over energy
shows that if the electron temperature is equal to the
Compton temperature 
 \begin{equation}
 T_{\rm C} \equiv
 \frac{ \langle\e^2 \rangle-\langle n\e^2 \rangle}
  {4 \langle \e \rangle} \;,
  \label{Tcompton}
 \end{equation}
the photon energy density remains constant although the
photon spectrum may evolve with time; here the average is
over photon energy, using the photon-number density
$N(\e) \equiv (J/\e)$ as the weighting function. Note
that, for a given photon spectrum, induced Compton
scattering {\it always decreases\/} the Compton
temperature.

\subsection{Implications of the Moment Equations for Static
Media}

As we mentioned in the Introduction, the Kompaneets
equation (\ref{Kompaneets}) is not strictly valid for
astrophysical systems, since it requires the radiation
field to be isotropic and hence that no radiation
leaves the system. If the radiation field is
anisotropic but there is no bulk motion, the system of
equations
 \begin{eqnarray}
 \partial_t J+\partial_i\Hr &=&
   \eta - \chi J\nonumber\\
 & & + n_e \sigma_T \left[\e\partial_\e
  \left(\frac{\e-4T_e}{m_e}\right)
 +\frac{T_e}{m_e}\e\partial_\e^2\e\right]J
   \label{0thstatic}
 \end{eqnarray}
 and
 \begin{eqnarray}
 \partial_t \Hr+\partial_j\Pr & = &
  -n_e\sigma_T\Hr-\chi\Hr\nonumber\\
 & &-\frac{2}{5}n_e\sigma_T\left[
\left(\frac{T_e-3\e}{m_e}\right)
 +\e\partial_\e\left(\frac{\e-4T_e}{m_e}\right)
 +\frac{T_e}{m_e}\e\partial_\e^2\e\right]\Hr
 \label{1ststatic}
 \end{eqnarray} 
 must be solved simultaneously together with closure
relations for $K^{ij}$ and $J$, which are usually
introduced as variable Eddington factors (see Mihalas
1978, p.~157). For simplicity, we have neglected the
effects of induced Compton scattering in equations
(\ref{0thstatic}) and (\ref{1ststatic}). The term  $(2
n_e\sigma_T/5)[(T_e-3\e)/m_e]H^i$ does not appear in the
first moment of the transfer equation derived by Chan
\& Jones (1975), Payne (1980), and Madej (1989, 1991),
because these authors did not use the appropriate
relativistic scattering cross section.

In a static medium and in the absence of absorption,
emission, and induced scattering, $F^i$, the radiation
force per unit volume on the electrons (see also Miller \&
Lamb 1995\markcite{ML1995}) can be obtained by integrating
equation (\ref{1ststatic}) over photon energy, which gives 
 \begin{equation}
 F^i= \frac{4\pi}{c} n_e \sigma_T \left( 1 -
   \frac{8}{5}\frac{\langle \e \rangle_{H^i}}{m_e}
   +2\frac{T_e}{m_e}\right)
   \int_0^\infty {H}^i\, d\e\;,
	\label{radforce}
 \end{equation}
 where $\langle \e \rangle_{H^i}$ is the average photon
energy, using $H^i$ as the weighting function. Blandford
\& Payne (1981a) neglected the terms in the square
brackets on the right side of equation (\ref{1ststatic})
and therefore the radiation force given by their
equation is incorrect to first order in $\e/m_e$ and
$T_e/m_e$. Equation~(\ref{radforce}) shows that the
volume radiation force on the electrons produced by
Compton scattering of the photons is different from that
obtained in the Thomson limit, and depends both on the
photon spectrum and on the electron temperature (see also
Fukue et al.\ 1985). Hence, to first order in $\e/m_e$
and $T_e/m_e$, the energy-integrated critical radial
radiation flux ${\cal F}^{\rm crit}$ that produces a
radially outward radiation force which exactly balances
the inward gravitational force of a massive object
depends on the photon spectrum and the electron
temperature. For a completely ionized hydrogen gas, the
critical radiation flux at radius $r$ is given
implicitly by the equation
 \begin{equation}
 {\cal F}^{\rm crit}=\frac{c m_p GM}{r^2 \sigma_T}
   \left(1 -\frac{8}{5}\frac{\langle \e
   \rangle_{{\cal F}^{\rm crit}_{\e}}}{m_e}
   +2\frac{T_e}{m_e}\right)^{-1}\;,
 \end{equation}
where $m_p$ is the proton mass, $M$ is the mass of the
object, and $\langle \e \rangle_{{\cal F}^{\rm
crit}_{\e}}$ is the average photon energy, using ${\cal
F}^{\rm crit}_{\e}$ as the weighting function. For
example, in the spectral formation region of many
neutron-star low-mass X-ray binaries, $\langle \e
\rangle_{{\cal F}^{\rm crit}_{\e}}$ is $\sim 1$~keV and
$T_e$ can be $\sim 25$~keV (Miller \& Lamb 1992; Psaltis
et al.\ 1995), in which case the critical flux is $\sim
10$\% smaller than the usual Eddington critical flux
computed assuming Thomson scattering. Note also that to
this order the critical luminosity $L_{\rm crit} \equiv
4\pi r^2 {\cal F}^{\rm crit}$ generally depends on
radius, because both $\langle \e \rangle_{{\cal F}^{\rm
crit}_{\e}}$ and $T_e$ generally depend on radius.

\subsection{Implications of the Moment Equations for Moving
Media}

Consider now a medium in which the electron bulk velocity
is not zero. Suppose first, for simplicity, that the
radiation field is isotropic in the {\it system\/} frame.
To make this situation concrete, consider a thought
experiment in which electrons are moving with uniform
and constant bulk velocity $\vec{V}$ through a box with
sides of length $L$. The electrons are assumed to be able
to pass through the walls of the box whereas photons
are confined inside the box and have a mean free path much
larger than $L$. Under these conditions, scattering of
photons by the walls of the box, which is much more
frequent than scattering of photons by the electrons,
keeps the photon distribution nearly isotropic.

In this situation the equation for the zeroth moment of
the specific intensity reduces to
 \begin{equation} 
 \partial_{t_c} J = 
 \frac{\e}{m_e}\partial_\e(\e J)+
 \left(\frac{T_e}{m_e}+\frac{V^2}{3}\right)
 \left(-4\e\partial_\e  
 +\e\partial_\e^2\e \right)J\;,
   \label{0thmomentisotropic}
 \end{equation}
 where for simplicity we have neglected absorption,
emission, and induced scattering.
Equation~(\ref{0thmomentisotropic}) shows that when the
radiation field is isotropic in the system frame,
Comptonization by the bulk motion of the electrons is
described entirely by terms that are second-order in $V$; all
terms that are first-order in $V$ vanish identically.

Suppose now that (i)~photons with energies $\e \ll T_e +
{\textstyle \frac{1}{3}} m_e V^2$ are injected into the box
with an isotropic momentum distribution, (ii)~the photons
are allowed to remain in the box for a distribution of
residence times that decreases exponentially with the
residence time, and (iii) 
 \begin{equation}
 y_b \equiv
 4\left(\frac{T_e}{m_e} +
 {\textstyle \frac{1}{3}}V^2 \right)
 \bar{t}_{\rm res}
 \lesssim 1 \;,
 \label{Comptony}
 \end{equation}
 where $\bar{t}_{\rm res}$ is the mean residence time
measured in units of the Compton time $(1/n_e\sigma_T)$. The
solution of equation (\ref{0thmomentisotropic}) is then a
power-law spectrum with a high-energy cutoff. At high and
low energies the spectrum is
 \begin{equation}
 J~=~\left\{
 \begin{array}{ll}
     \e^{3+\alpha}\;,
   & \e\ll T_e + {\textstyle \frac{1}{3}} m_e V^2\;,\\
     \e^3\exp[-\e/(T_e + {\textstyle \frac{1}{3}} V^2)]\;,
   & \e \gg  T_e + {\textstyle \frac{1}{3}} m_e V^2\;,
 \end{array} \right. 
 \end{equation}   
 where
 \begin{equation}
\alpha\equiv
    -\frac{3}{2}-\sqrt{\frac{9}{4}+\frac{4}{y_b}}\;.
 \end{equation}
 and the factor $\e^3$ arises because we are considering the
energy density rather than the photon occupation number.
This solution is a simple generalization of the solution to
the Kompaneets equation obtained by Shapiro, Lightman \&
Eardly\markcite{SLE1976} (1976; see also Rybicki \&
Lightman 1979, Ch.~7) and shows that {\it the terms in the
transfer equation that are second-order in the electron
bulk velocity produce a systematic increase in the energy
of the photons that is completely analogous to the
systematic increase in the energy of the photons produced
by the electron thermal motions}. In treating astrophysical
systems, the mean residence time  $t_{\rm res}$ in
equation~(\ref{Comptony}) is often expressed in terms of
$\tau^2$, where $\tau$ is the electron scattering optical
depth of the system. For example, if photons are injected
at the center of a uniform-density spherical cloud of
optical depth $\tau$, then the mean residence time is
$(3/\pi^2)(\tau + 2/3)^2$ (Sunyaev and
Titarchuk\markcite{ST1980} 1980).         

When the photon momentum distribution is not perfectly
isotropic, the characteristic time scales on which the
photon distribution changes because of systematic
downscattering and Comptonization by the electron
thermal and bulks motions are (see
eq.~[\ref{0thmoment}])
 \begin{eqnarray}
 t_{\rm down}^{-1} & \sim & n_e \sigma_T 
  \left(\frac{\e}{m_e}\right)\;,
  \label{downscatt}\\
 t_{\rm th}^{-1} & \sim & n_e \sigma_T 
  \left(\frac{4T_e}{m_e}\right)\;,
  \label{tthermal}\\
 t_{\rm V}^{-1} & \sim &  n_e \sigma_T 
  \vec{V}\cdot \left(\frac{\vec{H}}{J}\right)\;, 
  \label{tV1}\\
 t_{\rm V^2}^{-1} & \sim & n_e \sigma_T 
  \left(\frac{4V^2}{3}\right)\;, 
  \label{tV2}
 \end{eqnarray}
 where the last two time scales have been estimated
from the terms that are first- and second-order in
$V$.

Comparison of rates~(\ref{tthermal}) and (\ref{tV2}) shows
that Comptonization by the electron bulk motion is more
important than Comptonization by the electron thermal motion
if 
 \begin{equation}
 V^2 > \frac{3 T_e}{m_e}\;.
   \label{V2vsTe}
 \end{equation}
 In fact, Comptonization by the bulk motion
occurs  whether or not the radiation field is isotropic or the
bulk motion converges (see also
eq.~[\ref{0thmomentisotropic}]), contrary to the impression
given by Blandford \& Payne (1981a). Comparison of
rates~(\ref{tV1}) and (\ref{tV2}) shows that {\it if $J$
is sufficiently large compared to $H^i$, the effects of bulk
Comptonization described by the terms that are second-order
in $V$ are dominant compared to the effects described by the
terms that are first-order in $V$}.

In estimating the characteristic time scale for bulk
Comptonization, Blandford \& Payne (1981a; see also
Blandford \& Payne 1981b and Payne \& Blandford 1981)
used only one of the several terms in their equation
that are first-order in $V$, neglecting other terms that
generally also produce systematic upscattering or
downscattering of photons. As a result, the
characteristic time scale that they obtained for bulk
Comptonization is proportional to $(\nabla \cdot
\Vb)/3$, which they assumed to be proportional to $n_e
\sigma_T V^2/3$. This assumption is not generally valid.
When it is, their expression suggests that bulk
Comptonization is less important than thermal
Comptonization if $V^2 < 12T_e/m_e$.
However, comparison of rates~(\ref{tthermal})--(\ref{tV2})
shows that the bulk Comptonization terms that are second-order
in $V$ are already as important as the thermal Comptonization
terms when $V^2 \sim {3 T_e}/{m_e}$ and hence that these
terms can be more important than the thermal terms even if
$V^2 < 12T_e/m_e$.

\subsection{A Method of Solving the Transfer Equation for
Moving Media}

In more realistic models, the radiation field is not
isotropic in the system frame in the presence of bulk motion
(see, e.g., Miller \& Lamb 1993, 1996) because electron
scattering tends to isotropize the photon distribution in the
{\it fluid\/} frame. In this case, the full radiative
transfer equation must be solved. This can be done using the
method of variable Eddington factors (Mihalas 1980; see also
Mihalas 1978, pp.~201--203), in which the radiative transfer
equation and its zeroth and first moments are solved
iteratively. In this approach, the second and higher moments
of the specific intensity are related to the zeroth and first
moments via variable Eddington factors. The zeroth and first
moments of the radiative transfer equation are then solved
using initial guesses for the variable Eddington factors and
the source function is computed from the calculated moments
of the specific intensity. The radiative transfer equation is
then solved, the Eddington factors are updated, and the whole
procedure is repeated until convergence is achieved. 

A detailed study of the solutions of the equations derived
here for realistic models of astrophysical systems will
be reported elsewhere.

\bigskip It is a pleasure to thank G.\ Baym, V.\
Kalogera, J.\ Poutanen, L.\ Titarchuk, R.\ Turolla, L.\
Zampieri, and S.\ Zane for helpful discussions and comments
on the original manuscript. We are also grateful to V.\
Kalogera for help in checking the derivations. This research
was supported in part by NSF grant AST~93-15133 and NASA
grant NAG~5-2925.

%\newpage

\appendix

\begin{center}
{\bf APPENDICES}
\end{center}

\section{PHOTON KINETIC AND RADIATIVE TRANSFER EQUATIONS
\newline FOR SCATTERING BY AN ELECTRON FLUID\ \ \ \ \ \ }

In this appendix we give the photon kinetic equation and the
corresponding radiative transfer equation for scattering by an electron
gas with temperature $T_e$ and bulk velocity $\vec{V}$. 

We start from the photon kinetic equation (\ref{PKE}), transform all
the  quantities to the system frame using equations
(\ref{eo})--(\ref{lo}), and then average over the
electron velocity distribution. The resulting photon kinetic
equation is
\begin{eqnarray}
\frac{1}{n_e \sigma_T}\left(\partial_t+l^i\partial_i\right)\n & =&  
{\cal R}_1 \n  + \frac{3}{4}  \left( {\cal R}_2 \nj 
+{\cal R}_3^i\nf + {\cal R}_4^{ij} \npr + {\cal R}_5^{ijk} \nq +
{\cal R}_6^{ijkl} n^{ijkl}\right)\nonumber\\
& & +\frac{3}{2}\left(\frac{\e}{m_e}\right)\n \left(2+\e
   \partial_\e\right)
   \left(n-l^i \nf + l^i l^j \npr - l^i l^j l^k
   \nq\right), 
   \label{PKEave}
\end{eqnarray}
where
\begin{eqnarray}
{\cal R}_1 & = & -1 + 2 \frac{\e}{m_e} + \lub\;, \\
{\cal R}_2 & = &1 + 2 \frac{\e}{m_e} + \frac{\e}{m_e} \e \partial_\e 
   +2 \frac{T_e}{m_e} - \lub + \lub^2 -V^2 \nonumber\\
& & + \e \partial_\e \left[ 4 \frac{T_e}{m_e} - \lub + \lub^2\right]
    + \frac{1}{2}\e^2 \partial^2_\e \left[2
    \frac{T_e}{m_e}+\lub^2\right]\;,\\ 
{\cal R}_3^i & = & -2 \frac{\e}{m_e} l^i -\frac{\e}{m_e} \e \partial_\e
l^i- 4 \frac{T_e}{m_e} l^i + 2 V_i - 2 \lub l^i - 2\lub^2 l^i + 2 V^2
   l^i\nonumber \\
& & + \e \partial_\e \left[ -4 \frac{T_e}{m_e} l^i + V_i - 4 \lub V_i + 
   2 \lub^2 l^i\right] \nonumber\\
& & + \frac{1}{2}\e \partial^2_\e \left[-2 \frac{T_e}{m_e} l^i
   - 2 \lub V_i\right]\;,\\
{\cal R}_4^{ij} & = & l^i l^j + 2\frac{\e}{m_e} l^i l^j + \frac{\e}{m_e} 
   \e \partial_\e l^i l^j - 6 \frac{T_e}{m_e} l^i l^j 
   + \lub l^i l^j - 2 l^i V_j \nonumber\\
& & + \lub^2 l^i l^j - 8 \lub l^i V_j - 3 V^2 l^i l^j
   +4 V_i V_j\nonumber\\
& & +\e \partial_\e \left[ 4 \frac{T_e}{m_e} l^i l^j - \lub l^i l^j
   +  3 V_i V_j - \lub^2 l^i l^j \right]\nonumber\\
& & + \frac{1}{2} \e \partial^2_\e \left [2 \frac{T_e}{m_e} l^i l^j +
    V_i V_j
   + \lub^2 l^i l^j \right]\;,\\
{\cal R}_5^{ijk} & = & -2 \frac{\e}{m_e} l^i l^j l^k - \frac{\e}{m_e} \e
\partial_\e l^i l^j l^k + 4\frac{T_e}{m_e} l^i l^j l^k \nonumber\\
& & + 4 l^i l^j V_k - 8 \l^i V_j V_k +4 \lub l^i l^j V_k \nonumber\\
& & + \e \partial_\e \left[-4 \frac{T_e}{m_e} l^i l^j l^k + l^i l^j V_k 
   - 4 \lub l^i l^j V_k - 2 l^i V_j V_k\right]\nonumber\\
& & + \frac{1}{2}\e^2 \partial^2_\e \left[-2 \frac{T_e}{m_e}l^i l^j l^k
   - 2  \lub l^i l^j V_k\right]\;,\\
{\cal R}_6^{ijkl} & = & 10 l^i l^j V_k V_l + 5 \e \partial_\e l^i l^j
V_k V_l  + \frac{1}{2}\e^2 \partial^2_\e l^i l^j V_k V_l\;.
\end{eqnarray}
The corresponding radiative transfer equation is
\begin{eqnarray}
\frac{1}{n_e \sigma_T}\left(\partial_t
+l^i\partial_i\right)I(\l,\e) &=& 
{\cal L}_1 I(\l,\e) +\frac{3}{4}  \left( {\cal L}_2 J  +{\cal
L}_3^i\Hr + {\cal L}_4^{ij} \Pr + {\cal L}_5^{ijk} \Qr +
{\cal L}_6^{ijkl} R^{ijkl}\right)\nonumber\\ 
& & +\frac{3}{4\e^3}\left(\frac{\e}{m_e}\right)I(\l,\e)
  \left(\e\partial_\e-1\right)
   \left(J-l^i \Hr + l^i l^j \Pr - l^i l^j l^k \Qr\right)\;, 
   \label{RTEave}
\end{eqnarray}
where
\begin{eqnarray}
{\cal L}_1 & = & -1 + 2 \frac{\e}{m_e} + \lub\;, \\
{\cal L}_2 & = &1 - \frac{\e}{m_e} + \frac{\e}{m_e} \e \partial_\e 
   +2 \frac{T_e}{m_e} +2 \lub + 4\lub^2 -V^2 \nonumber\\
& & + \e \partial_\e \left[ -2 \frac{T_e}{m_e} - \lub -2 \lub^2\right]
    + \frac{1}{2}\e^2 \partial^2_\e \left[2
    \frac{T_e}{m_e}+\lub^2\right]\;,\\ 
{\cal L}_3^i & = & \frac{\e}{m_e} l^i -\frac{\e}{m_e} \e \partial_\e
l^i- 4 \frac{T_e}{m_e} l^i - V_i - 2 \lub l^i - 8\lub^2 l^i + 2 V^2
   l^i\nonumber \\
& & + \e \partial_\e \left[ 2 \frac{T_e}{m_e} l^i + V_i +2 \lub V_i + 
   2 \lub^2 l^i\right] \nonumber\\
& & + \frac{1}{2}\e \partial^2_\e \left[-2 \frac{T_e}{m_e} l^i
   - 2 \lub V_i\right]\;,\\
{\cal L}_4^{ij} & = & l^i l^j -\frac{\e}{m_e} l^i l^j + \frac{\e}{m_e} 
   \e \partial_\e l^i l^j - 6 \frac{T_e}{m_e} l^i l^j 
   + 4\lub l^i l^j - 2 l^i V_j\nonumber\\
& & + 10\lub^2 l^i l^j - 8 \lub l^i V_j - 3 V^2 l^i l^j 
   + V_i V_j\nonumber\\
& & +\e \partial_\e \left[ -2 \frac{T_e}{m_e} l^i l^j - \lub l^i l^j
   - 4\lub^2 l^i l^j \right]\nonumber\\
& & + \frac{1}{2} \e \partial^2_\e \left [2 \frac{T_e}{m_e} l^i l^j 
   + V_i V_j+ \lub^2 l^i l^j \right]\;,\\
{\cal L}_5^{ijk} & = & \frac{\e}{m_e} l^i l^j l^k - \frac{\e}{m_e} \e
\partial_\e l^i l^j l^k + 4\frac{T_e}{m_e} l^i l^j l^k \nonumber\\
& & + l^i l^j V_k - 2 l^i V_j V_k +4 \lub l^i l^j V_k
   \nonumber\\
 & & + \e \partial_\e \left[2 \frac{T_e}{m_e} l^i
l^j l^k + l^i l^j V_k 
   +2 \lub l^i l^j V_k - 2 l^i V_j V_k\right]\nonumber\\
& & + \frac{1}{2}\e^2 \partial^2_\e \left[-2 \frac{T_e}{m_e}l^i l^j l^k
   - 2  \lub l^i l^j V_k\right]\;,\\
{\cal L}_6^{ijkl} & = & l^i l^j V_k V_l + 2 \e \partial_\e l^i l^j V_k
V_l  + \frac{1}{2}\e^2 \partial^2_\e l^i l^j V_k V_l\;. 
\end{eqnarray}

\section{EMISSION AND ABSORPTION PROCESSES
\newline IN THE SYSTEM FRAME \ \ \ \ \ \ \ \ \ \ \ \ \ }

In the absence of scattering, the transfer equation in the
system frame becomes simply (see, e.g., Mihalas \& Mihalas
\markcite{MM1} 1984, p. 422)
 \begin{equation}
 \left(\partial_t + l^i \partial_i\right) I(\l,\e) = 
   \left(\frac{\e}{\e_f}\right)^2 \eta(\e_f) -
  \left(\frac{\e_f}{\e}\right)\chi(\e_f) I(\l,\e),
 \label{RTEabsemis}
 \end{equation}
where again the subscript `f' refers to quantities in the
fluid frame and we have for simplicity assumed that the
emission coefficient $\eta$ and the absorption coefficient
$\chi$ are isotropic in this frame. In writing equation
(\ref{RTEabsemis}) we have also used the fact that the
absorption and emission coefficients are defined in the fluid
frame and that
$(\eta/\e^2)$ and
$(\e\chi)$ are Lorentz invariants.

We can now expand $\eta(\e_f)$ to second order in $\vec{V}$
as
 \begin{eqnarray}
\eta(\e_f) & = &\eta[\e+(\e_f-\e)]\nonumber\\
& \simeq & \eta_\e+(\e_f-\e)\partial_\e\eta_\e
+\frac{1}{2}(\e_f-\e)^2\partial^2_\e\eta_\e
\nonumber\\
& \simeq & \eta_\e+\left(\frac{1}{2}V^2-\hat{l}\cdot\Vb\right)
 \e\partial_\e\eta_\e+
 \left(\hat{l}\cdot\Vb\right)^2
  \e^2\partial^2_\e\eta_\e\;,
 \end{eqnarray}
 where $\eta_\e=\eta(\e)$. We also expand $\chi(\e_f)$ in
a similar way. The transfer equation to second order in
$V$ then becomes
 \begin{eqnarray}
\left(\partial_t +l^i \partial_i \right)I(\l,\e) & = &
   \left\{\left[1+2(\l\cdot\Vb)+3(\l\cdot\Vb)^2-V^2
   \right]\right.\nonumber\\
& & \;\;\;\;\;\;\;\;\;\; \left.
   +\left[-(\l\cdot\Vb)-2(\l\cdot\Vb)^2+\frac{1}{2}V^2
   \right]\e\partial_\e
   +\frac{1}{2}(\l\cdot\Vb)^2\e^2\partial^2_\e\right\}
   \eta_\e\nonumber\\
& & -\left\{\left[1-(\l\cdot\Vb)+\frac{1}{2}V^2\right] 
  +\left[-(\l\cdot\Vb)+(\l\cdot\Vb)^2+\frac{1}{2}V^2\right]
   \e\partial_\e\right. \nonumber\\
& &  \;\;\;\;\;\;\;\;\;\; \left.
    +\frac{1}{2}(\l\cdot\Vb)^2
   \e\partial_\e\right\}\chi_\e I(\l,\e)\;.
  \label{RTEabs}
 \end{eqnarray} 
 We can then obtain the zeroth and first moments of the
transfer equation,
 \begin{eqnarray}
\partial_t J+\partial_i H^i & = &
   \eta_\e - \chi_\e J + V^i H^i 
   \left[\chi_\e+\e\partial_\e\chi_\e\right]\nonumber\\
& &-\frac{1}{6}V^2 \e\partial_\e\eta_\e
   +\frac{1}{6}V^2 \e^2\partial^2_\e\eta_\e
   -\frac{1}{2}\chi_\e J V^2 - \left[\frac{1}{2}J V^2 +
  \Pr V_i V_j\right]
  \e\partial_\e\chi_\e\nonumber\\
& &-\frac{1}{2}\Pr V_i V_j\e^2\partial^2_\e\chi_\e
  \label{0thabs}
 \end{eqnarray}
and
 \begin{eqnarray}
 \partial_t H^i+\partial_j K^{ij}
 & = & -\chi_\e H^i +\frac{1}{3} V^i \left(2 \eta_\e -
  \e\partial_\e\eta_\e\right)+
  \Pr V_j \left(\chi_\e +
  \e\partial_\e\chi_\e\right)\nonumber\\
& & - \frac{1}{2}\chi_\e H^i V^2 - \left(\frac{1}{2} H^i V^2 
  + \Qr V_j V_k\right)
   \e\partial_\e\chi_\e
  -\frac{1}{2}\Qr V_j V_k
   \e^2\partial^2_\e\chi_\e\;.
   \label{1stabs}
 \end{eqnarray}

Blandford \& Payne (1981a) added a photon source term to
the right side of their transfer equation, without adding
any corresponding absorption term. This is fundamentally
inconsistent with thermodynamics (if absorption is not
included, the photon source can never come into
thermodynamic equilibrium with the radiation field). It
is equivalent to including the terms on the right sides of
equations~(\ref{RTEabs})--(\ref{1stabs}) that involve the
emission coefficient $\eta_\e$ but neglecting all the
terms that involve the absorption coefficient $\chi_\e$.

Neglecting the absorption terms compared to the emission
term is a valid {\it approximation\/} only if the
specific intensity of the radiation field is negligible
compared to the source function $\cal{S}_\e =
\eta_\e/\chi_\e$ {\it at all photon energies}, i.e., only
if self absorption is never important at any energy, anywhere
in the system. This is rarely the case in astrophysical
systems. For example, it is not the case in the Comptonizing
regions around accreting neutron star and black hole
X-ray sources.

Neglecting the absorption terms in the transfer equation
leads to equations that have a different mathematical
character from equations (\ref{RTEabs})--(\ref{1stabs}),
because in this case the right sides of the transfer and
moment equations do not depend explicitly on the
radiation field.

\end{document}